\documentclass{sig-alternate-05-2015}

\begin{document}

\setcopyright{acmcopyright}

\doi{10.475/123_4}

\isbn{xxx-xxx}

\conferenceinfo{KDD '16}{San Francisco, CA}

\acmPrice{\$15.00}

\title{The Myopia of Crowds:\\ A Study of Collective Evaluation on Stack Exchange}

\numberofauthors{5} 
\author{
\alignauthor
Keith Burghardt\\
       \affaddr{Dept of Physics}\\
       \affaddr{University of Maryland}\\
       \affaddr{College Park, MD, USA}\\
       \email{keith@umd.edu}
\alignauthor
Emanuel F. Alsina\\
       \affaddr{Dept of Physics, Mathematics and Informatics}\\
       \affaddr{U. Modena and Reggio Emilia}\\
       \affaddr{Modena, Italy}\\
       \email{\small emanuelfederico.alsina@unimore.it}
 \alignauthor
Michelle  Girvan\\
       \affaddr{Dept of Physics}\\
       \affaddr{University of Maryland}\\
       \affaddr{College Park, MD, USA}\\
       \email{girvan@umd.edu}
\and
 \alignauthor
William Rand\\
       \affaddr{Ctr for Complexity in Business}\\
       \affaddr{University of Maryland}\\
       \affaddr{College Park, MD, USA}\\
       \email{wrand@umd.edu}
 \alignauthor
Kristina Lerman\\
       \affaddr{Information Sciences Institute}\\
       \affaddr{U. Southern California}\\
       \affaddr{Marina del Rey, CA, USA}\\
       \email{lerman@isi.edu}
}
\newcommand\qa{Q\&A}
\newcommand{\note}[1]{{\color{red} #1}}

\maketitle
\begin{abstract}
Crowds can often make better decisions than individuals or small groups of experts by leveraging their ability to aggregate diverse information. Question answering sites, such as Stack Exchange, rely on the ``wisdom of crowds'' effect to identify the best answers to questions asked by users. We analyze data from 250 communities on the Stack Exchange network to pinpoint factors affecting which answers are chosen as the best answers. Our results suggest that, rather than evaluate all available answers to a question, users rely on simple cognitive heuristics to choose an answer to vote for or accept. These cognitive heuristics are linked to an answer's salience, such as the order in which it is listed and how much screen space it occupies. While askers appear to depend more on heuristics, compared to voting users, when choosing an answer to accept as the most helpful one, voters use acceptance itself as a heuristic: they are more likely to choose the answer after it is accepted than before that very same answer was accepted. These heuristics become more important in explaining and predicting behavior as the number of available answers increases. Our findings suggest that crowd judgments may become less reliable as the number of answers grow.
\end{abstract}


\section{Introduction}



Are crowds wiser than informed individuals? 
Generally speaking, a crowd's collective opinion---whether through votes, likes, or thumbs up/down---is often used to rank order items in crowdsourcing systems, which determines how much attention they receive~\cite{ghosh2014game}, as well as users' incentives for participating~\cite{jain2009designing}. The assumption is that collective opinions outperform individual experts, an observation long seen in a variety of contexts~\cite{surowiecki2005wisdom,Wagner10,Chen14}, even when they are less-informed than the experts. Recent evidence, however, has shown that the collective decision of the crowd is not foolproof. One known limitation, for example, is social influence, which biases individual judgments and  degrades crowd performance~\cite{Lorenz11}, obscuring the underlying quality of choices \cite{Salganik06}. We try to answer whether crowd wisdom limitations affect a common crowdsourcing application, question answering boards.

We carry out an empirical study of Stack Exchange\footnote{
{http://stackexchange.com}}, a network of more than a hundred question answering (\qa) communities,  where millions of people post questions on a variety of topics, and others answer them asynchronously.
Like other \qa~sites, such as Quora and Yahoo! Answers, Stack Exchange has a number of features for enhancing collaborative knowledge creation. In addition to asking and answering questions, users can evaluate answers by (1) \emph{voting} for them, and (2) askers can \emph{accept} a specific answer to their question. The votes, in aggregate, reflect the crowd's opinion about the quality of content, and are used by Stack Exchange to surface the right answers.
They also provide a lasting value to the community~\cite{Anderson12}, enabling future users to identify the most helpful answers to questions without asking the questions themselves.

We find that the number of answers users parse through can dramatically affect how users choose answers, including a greater reliance on heuristic-like answer attributes, potentially limiting the usefulness of question answering boards. In addition, we find behavior biases allow for users to choose answers in an increasingly predictable way, as the number of answers increases, running counter to our intuition that increasing the numbers of choices makes user decisions less predictable.

Alternatively, work also addresses some of the challenges of data heterogeneity. Large-scale datasets of human behavior, such as this one, provide new opportunities to study decision-making processes in crowdsourcing systems. In contrast to laboratory studies, which typically involve dozens of subjects, behavioral data are collected from millions of people under real-world conditions. Mining observational behavioral data, however, presents significant computational and analytic challenges. Human behavior is noisy and highly heterogeneous: aggregating data to improve the signal-to-noise ratio may obscure underlying patterns in heterogeneous data and even lead to nonsensical conclusions about human behavior~\cite{Vaupel85heterogeneity}. We discover that splitting data by the number of answers addresses one of the larger sources of user heterogeity, potentially providing greater predictive power in future models.

\textbf{Our Contributions.}

We use penalized regression to uncover factors associated with users' decisions to vote for or accept answers on all Stack Exchange communities. To partly control for heterogeneity, we split data by community type (technical, non-technical, meta) and leave out the largest community to check the robustness of results. In all cases, behavior was qualitatively the same and quantitatively similar. We find that the a significant source of behavioral heterogeneity is the number of existing answers to questions. To account for this, we separate data according to the number of answers questions have at the time that a user makes a decision about which answer to vote for (or accept).

We find that a few answer attributes are important in our regressions, including the order in which the answer appears, its share of words compared to the other available answers to the question, and whether it was accepted by the asker. This appears to imply that users rely on simple heuristics to choose an answer based on its rank, how much screen space it occupies, or whether it was approved by others. These heuristics may be useful proxies for answer quality, but our work suggests otherwise. For example, voters are more likely to choose an accepted answer after it has been accepted than before. Although answer acceptance is often viewed as a standard of answer quality~\cite{Shah10,Kim09,Agichtein08}, the only discernable difference in an answer after acceptance is a signal that the asker chose this answer, suggesting users view acceptance as a useful signal about quality, but are less able to discern that quality on their own.

We also find that heuristics better explain user behavior as the number of available answers to a question grows. Two different explanations are feasible.
First, as the number of answers to a question grows, users may become less willing to thoroughly evaluate all answers, instead increasingly relying on cognitive heuristics when choosing an answer. A similar effect exists in other domains. For instance, information overload impacts consumer's choice of products~\cite{scheibehenne2010can} and the spread information in online social networks~\cite{Hodas14srep,Rodriguez14}.
An alternative explanation is that later voters are different and happen to rely more on cognitive heuristics compared to people who vote early. This view is potentially supported by the observation that users who answer early in a question's life cycle on Stack Overflow, a programming-related community on Stack Exchange, have higher reputation than users who answer later~\cite{Anderson12}; therefore, time acts as a potential source of heterogeneity. In either case, the finding that voters rely more on heuristics as the number of answers grows points to a limitation of the ``wisdom of crowds'' effect on Stack Exchange: crowd's judgments become less reliable as proxies of quality as questions accumulate answers.

The rest of the paper is as follows. In the related work section, we review work related to our current analysis, while, in the materials and methods section, we discuss our data and ways in which we analyze it. Next, in the results section, we discuss our main findings. Finally, in the conclusion section, we review our findings, discuss future work, and discuss ways to improve upon question answering sites.

\section*{Related Work}
\label{sec:related}

Prior research on \qa~sites has shown that a variety of attributes can provide useful insights into content quality \cite{Agichtein08, blooma, wang, sakai}. For example, Kim and Oh \cite{Kim09} examined how users evaluate information in Yahoo! Answers forums, by examining the comments askers leave on answers. They found socioemotional-, content-, and utility-related criteria are dominant in the choice of the best answer, and found users evaluate information based not only upon the content, but also on cognitive and collaborative aspects. Adamic et al. \cite{adamic} conducted a large-scale network analysis of Yahoo! Answers, trying to predict which answers would be judged best and found that, for both technical and non-technical sites, answer length and the number of other answers the asker has to choose from are the most significant features to predict the future best answer. A preference for longer answers, however, diminishes with the number of answers. One limitation in these previous studies, however, is in assuming that the answer an asker chose was the ``best'' answer, and did not correct for asker biases when choosing any answer.

Several authors \cite{Shah10,Kim09,Agichtein08} used logistic regression to determine which attributes best describe high quality answers, although, again, it is often assumed that a``high quality" answer is one an asker accepts, a conclusion that our work casts doubt on.
Other works have examined the impact of answer order on answer quality. Anderson et al. \cite{Anderson12} found that early answers in Stack Overflow (the Stack Exchange community that deals with programming questions) tend to be posted by expert users with higher reputation, and subsequent answers come from lower reputation users. While the first answer tends to be more appreciated by the asker, the longer a question goes unanswered, the less likely that an answer will eventually be accepted. Similarly, Rechavi and Rafaeli \cite{rechavi} concluded that askers use response time as a parameter at evaluation time. However, this hypothesis was refuted in other works. Shah \cite{shah} analyzed the responsiveness in Yahoo! Answers forums, finding that more than 90\% of the questions receives an answer within an hour. However, satisfactory answers may take longer, depending on the difficulty of the questions. Interestingly, our work, discussed in the Results section, suggests that answer age and chronological order are not particularly important attributes for askers or voters. In part this is because high reputations answerers do not strongly affect whether an answer gets voted on (not shown). Older answers, however will accumulate more votes and will therefore be more likely to be voted on
, but the main driver appears to by answer attributes not directly dependent on time.

Unlike previous studies, we examine how voting may be affected by various answer attributes. This is an an important area to study, because people often use votes as a signal of the best answer to a particular problem. One previous study that also attempted to tackle this problem deduced a set of possible factors that indicate bias in user voting behavior~\cite{chen}. They provided a method to calibrate the votes inside \qa~sites, principally based on the average value of the answer and the average vote value received in the answerer history. This type of calibration is useful to restrict the effects of users who are trying to game the system, or to signal the reputation of answerers. Our work, however, answers a different set of questions: we want to find the role heuristics play in answer evaluation, how voter and asker behaviors differ, and what drives heterogeneity within voter and asker populations.
The role of heuristics in human decisions has been studied by behavioral economics~\cite{Kahneman03, simon1982models}, but, 
to the best of our knowledge, our work is the first that investigates the potential impact of heuristics on the performance of crowdsourcing systems.

\section{Data and Methods}\label{Methods}
Stack Exchange launched in 2008 with Stack Overflow, its first \qa~community for computer programming questions. Over time, Stack Exchange has added more communities covering diverse topics: 
\begin{itemize}
	\item 49 \textit{Technical} communities on topics, such as Programming, Server Faults, Information Security, Apple, Android and Ubuntu;
	\item 33 \textit{Culture and recreation} communities, e.g., English Language Learners, Bicycles, Videogamers Platforms, and Anime \& Manga;
	\item 17 \textit{Life and Arts} communities on topics related to the everyday life: e.g.,  Cooking, Photography, DIYers, and Movies \& TV;
	\item 16 \textit{Science} communities, e.g., Mathematics, Statistics, Biology, and Philosophy;
	\item 4 \textit{Business} communities on topics, such as Bitcoin, Project Management, and Finance;
\end{itemize}
There is a \emph{meta} board for each community where users discuss the workings and policies of the community: e.g., in Meta Stack Overflow users discuss the policies of Stack Overflow rather than computer programming itself. Posts that are overly subjective, argumentative, or likely to generate discussion rather than answers, are removed from the website. 

\begin{figure}[th]
	\centering
 		\includegraphics[width=1\columnwidth]{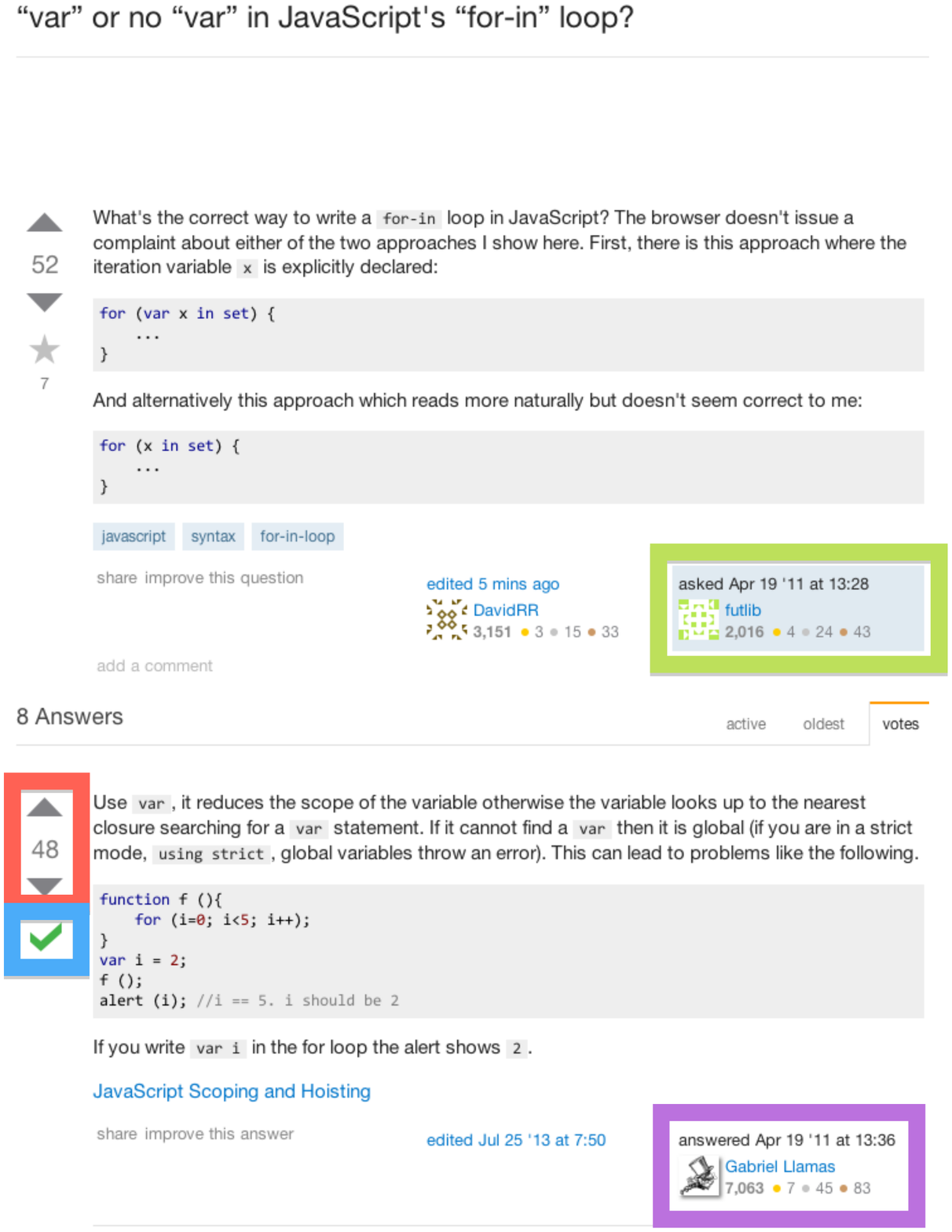}
	\caption{
	A screenshot of a Stack Exchange web page, showing a question (at top) and answers listed below in default order. The score next to the answer (red box), is defined by upvotes minus downvotes, and the green checkmark (blue box) denotes that the answer was accepted by the asker. We also consider other factors, including the times the question was asked (green box) and the answer was provided, as well as the answerer's reputation (purple box).
}
	\label{fig:screenshot}
\end{figure}

A user can post a question, which may receive multiple answers from different people, as shown in Fig.~\ref{fig:screenshot}. The asker can \emph{accept} an answer, which generally signifies that the asker finds it helpful. Regardless of acceptance, others can \emph{vote} an answer up (or down) if they think that it provides helpful (or irrelevant) information. By upvoting more helpful answers, a community collectively curates the information for both askers and future users interested in the same topic. The difference between the up and down votes is the \emph{score} of the answer. Answers with higher scores are shown at the top of the list of answers to the question, so that they are easier to find (answers with the same score are shown in random order). Figure \ref{fig:screenshot} shows an example question with answers, score for both answers and question,
the time the answer was submitted, answerer's reputation, and whether the answer was accepted by the asker.

For our study we used anonymized data consisting of all user contributions to Stack Exchange from 2009 until September 2014\footnote{
{https://archive.org/details/stackexchange}}. The data contain information about
five million posts (questions and answers) and 23 million votes. In particular, we used the data of 250 communities, 
including information related to the posts: the ID of the post, creation date, type of post (question or answer), ID of the relative question (in case of answers), the ID of the eventually accepted answer (in case of questions), and the content of the post; and related to the history of the votes made on each single post: the type of vote (up, down votes, or acceptance), the ID of the related post, and the time of assignment. In addition, we considered the information related to the users, such as the ID of the user, creation date (date of the sign up), and reputation. Particularly, we calculated the reputation of the users at the moment they asked or answered a question, considering the rules of Stack Exchange\footnote{
{http://meta.stackexchange.com/how-does-reputation-work}}.
\begin{figure}[tbh]

%
%
\centering
\begin{tabular}{@{}c@{}c@{}c@{}}
		\includegraphics[width=0.33\columnwidth]{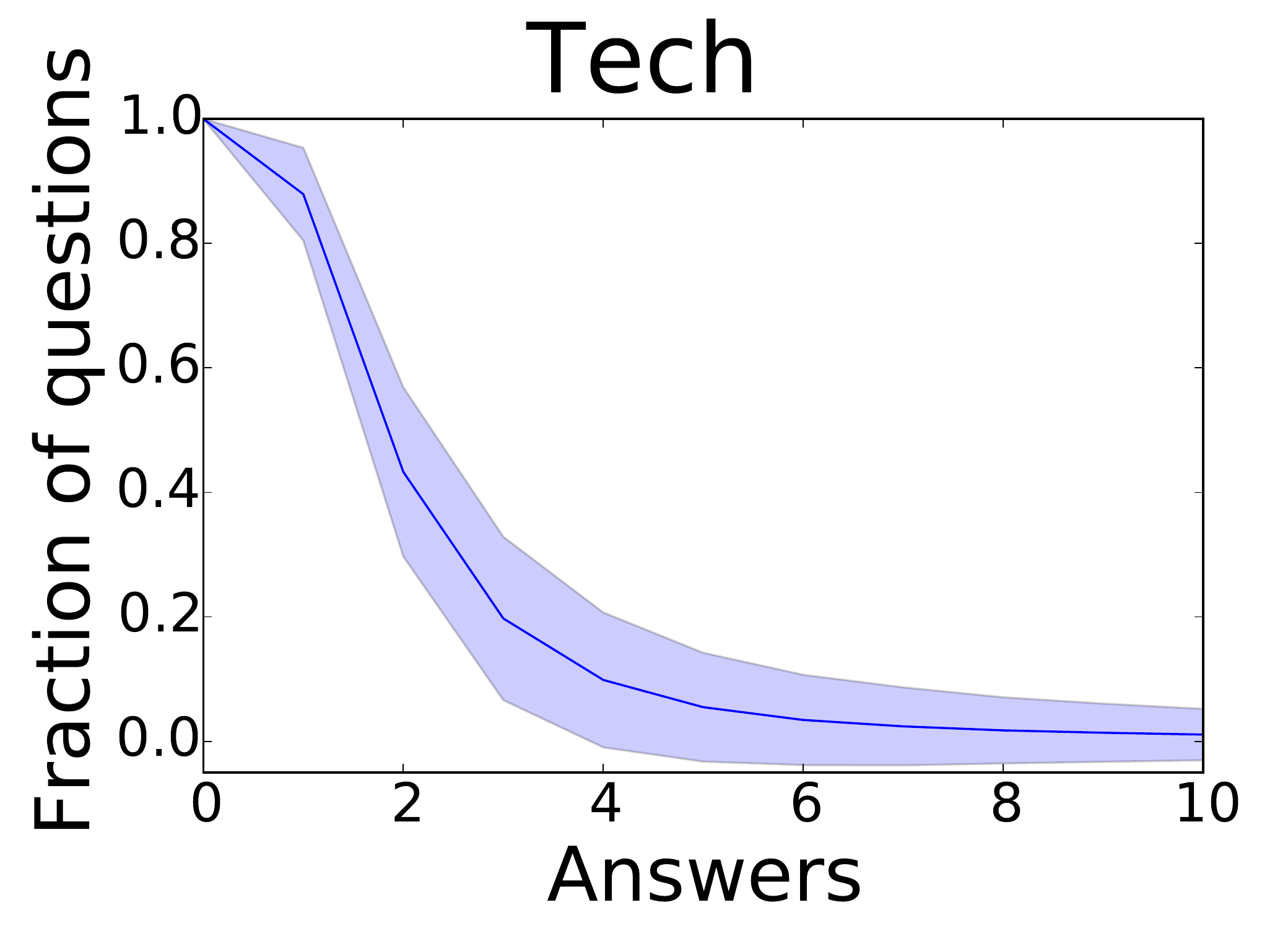}
&
	\includegraphics[width=0.33\columnwidth]{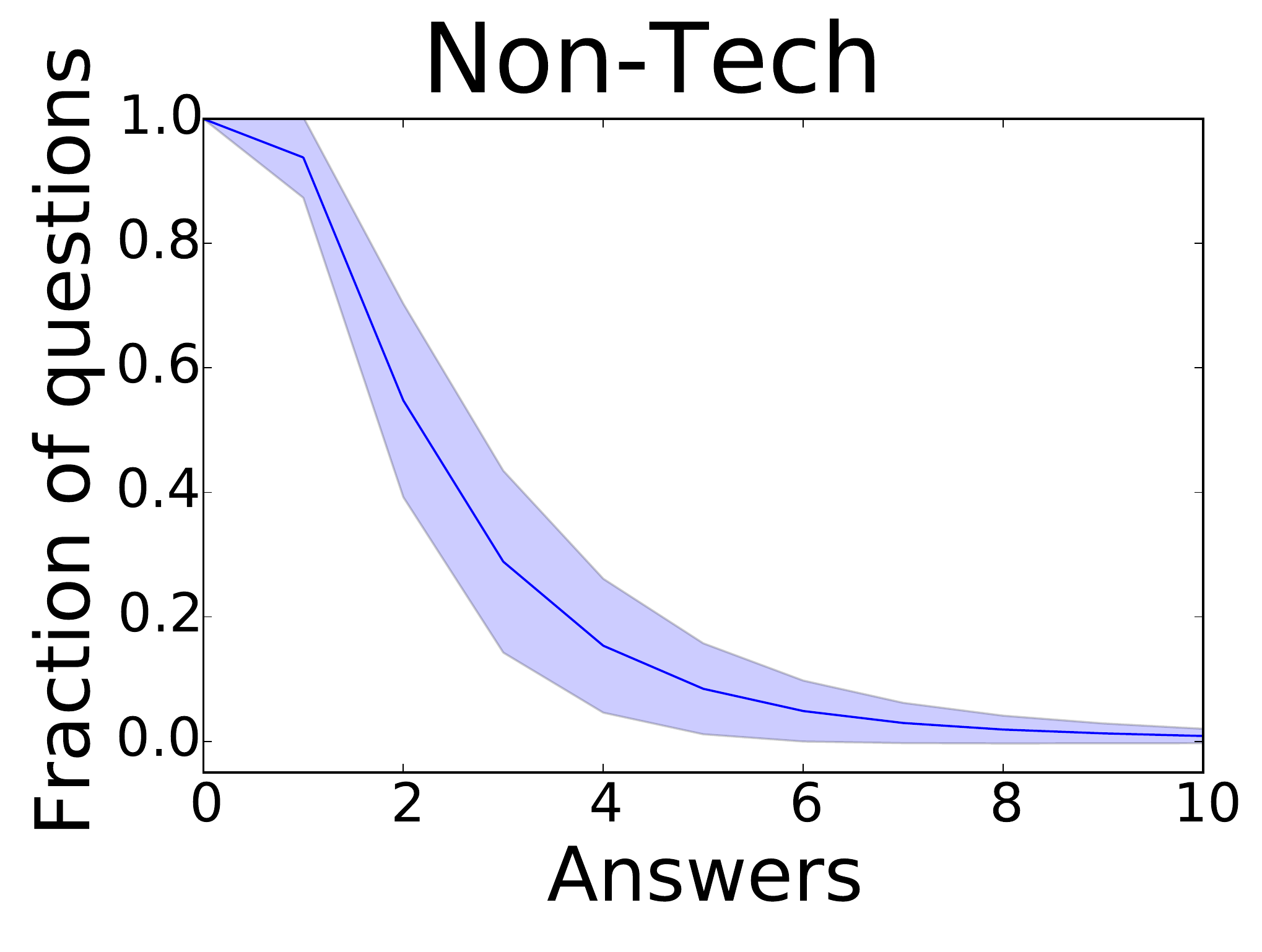}
&
	\includegraphics[width=0.33\columnwidth]{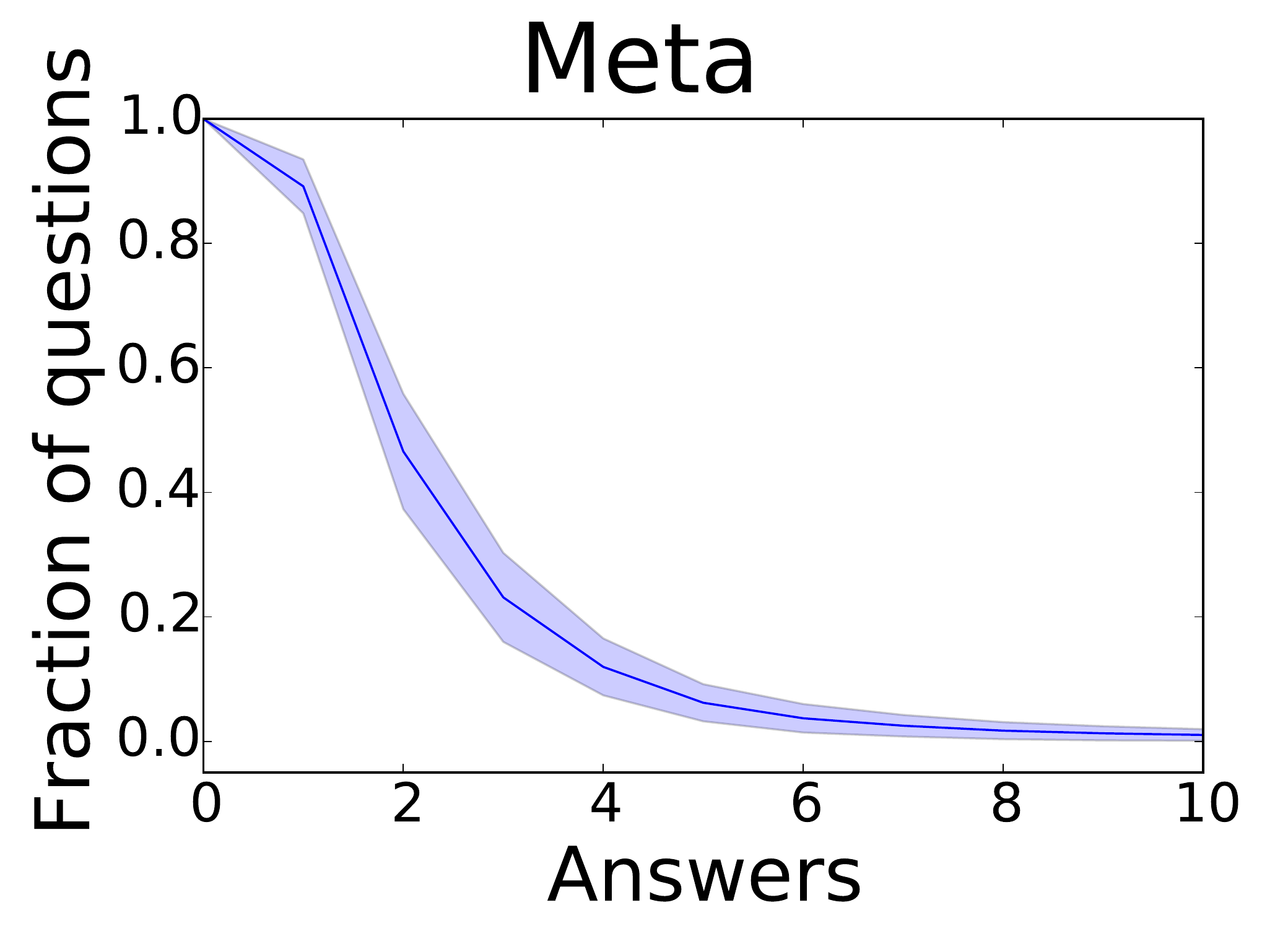}
\\
	\includegraphics[width=0.32\columnwidth]{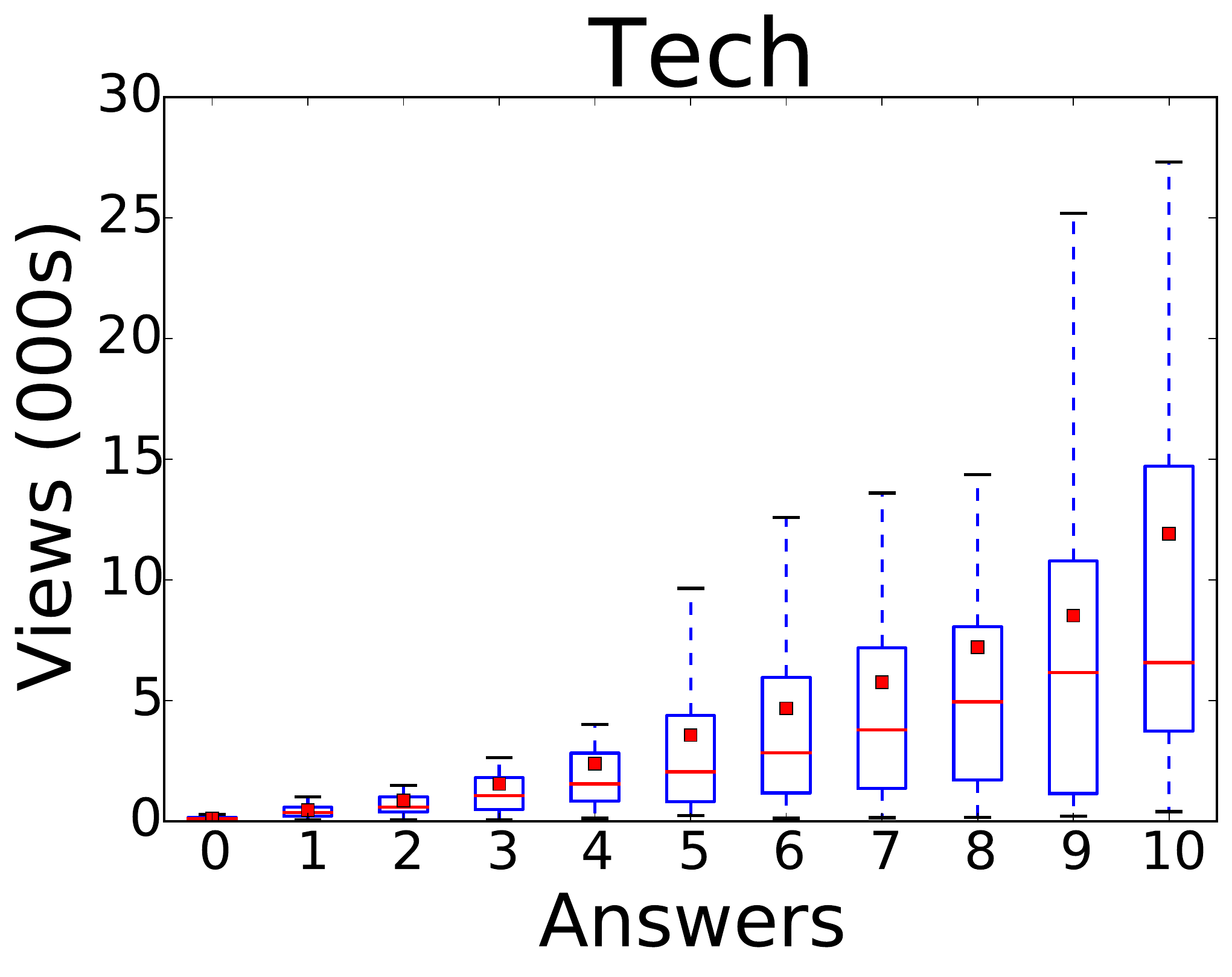}
&
	\includegraphics[width=0.32\columnwidth]{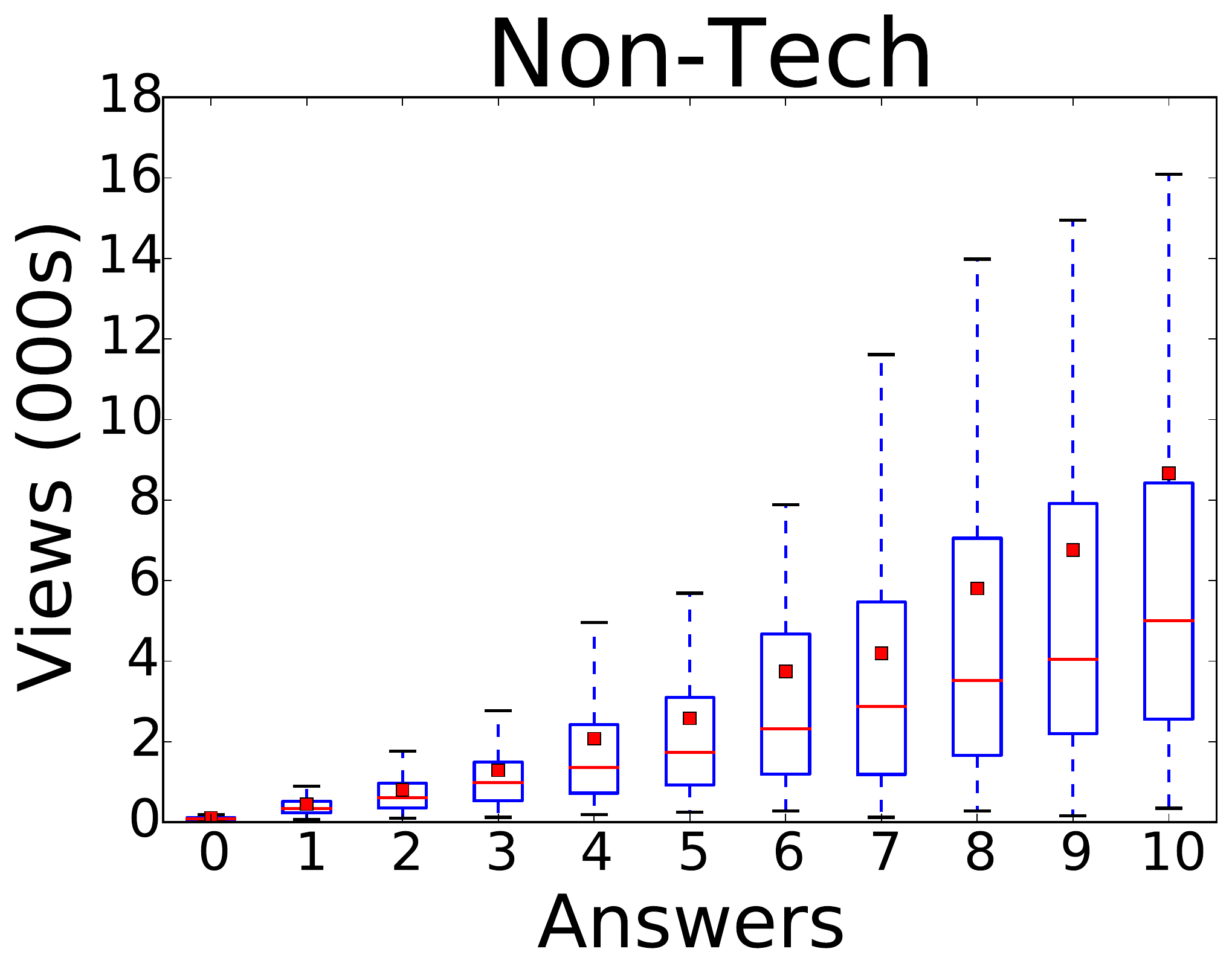}
&
	\includegraphics[width=0.32\columnwidth]{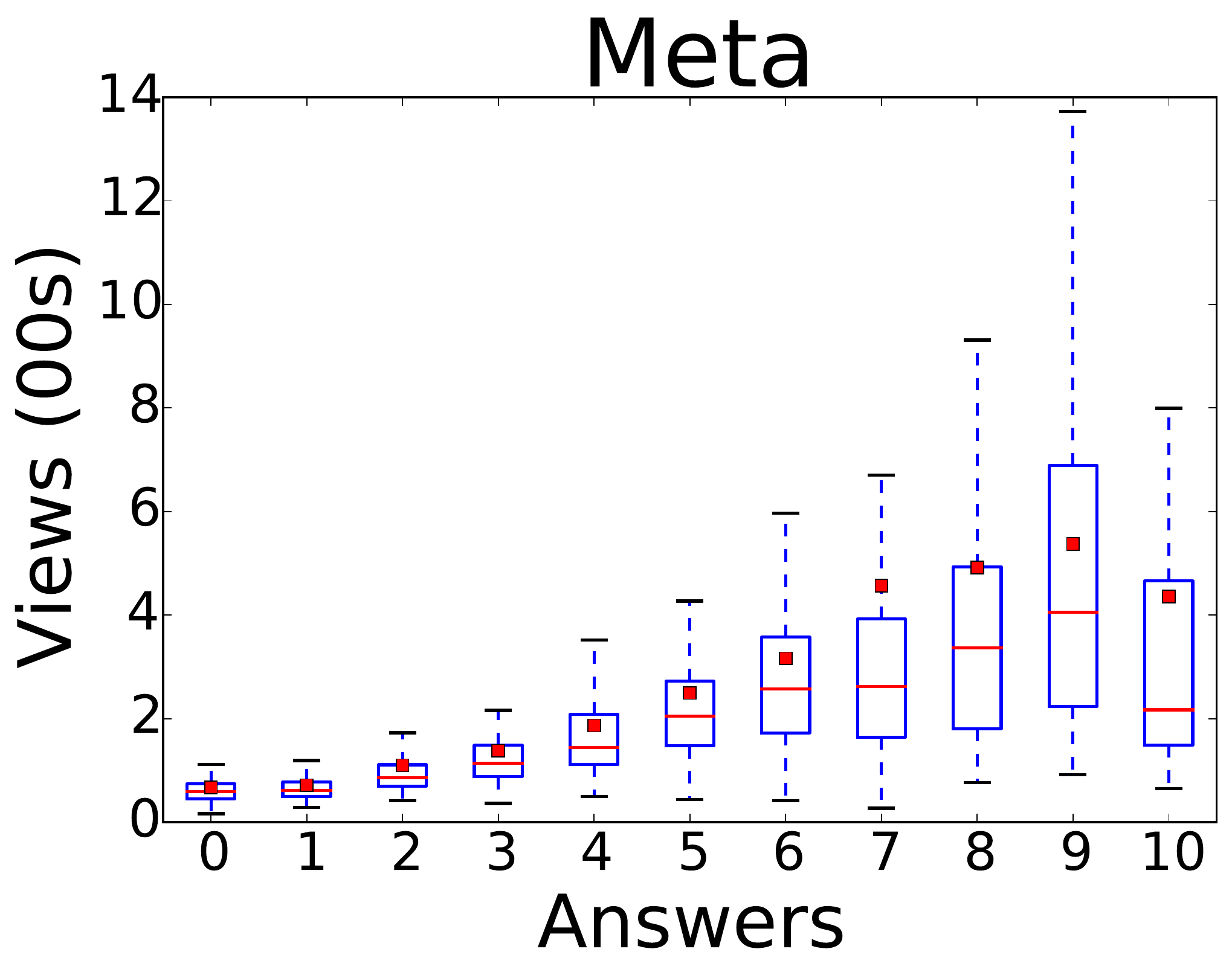}
\end{tabular}
\caption{(Top row) Complementary cumulative distribution of the final number of answers posted in reply to a question as of August, 2014, on (a) technical, (b) non-technical, and (c) meta sites. Shaded areas correspond to the standard deviation in the distributions. (Bottom row) Number of views per question (in August, 2014) as a function of the number of answers on (d) technical, (e) non-technical, and (f) meta sites. Boxes indicate 50\% confidence intervals, with a red line to indicate the median view count, and a red dot to represent the mean viewcount.}
\label{fig:NumbAnswers}
\end{figure}

Each question on these communities received almost three answers, on average. The ``Programming Puzzles \& Code Golf'' community had the highest number of average answers per question at $8$, while the ``Magento'' site had an average of only $1$.
About 10\% of the questions went unanswered, and 42\% received only one answer. Only the questions that received two or more answers were included in our study.
Figure \ref{fig:NumbAnswers} shows the complementary cumulative distribution of the number of answers posted for each question on technical, non-technical, and meta communities.
We considered an average of 11k answers per community, although this varied significantly. Technical communities had on average twice as many answers as non-technical communities, and an order of magnitude more than meta communities, which were not broad in appeal. Askers accepted an answer 59\% of the time in technical communities, and 59\% in non-technical communities,
but, curiously, only 37\% of the questions in meta boards were similarly accepted.
Answers with votes consist of 86\% (std 7\%) of the total, but this too varied across site types. 78\% of the answers on technical sites have votes, versus 86\% in non-technical and 88\% in meta communities (all differences are statistically significant with $p< 10^{-3}$ using t-tests).
The median time to obtain the first answer was 2.77 hours, and the eventually accepted answer, 4.47 hours.
As the community matures, the questions become more complex, which attracts the attention of users who may focus on different facets of the problem, posting multiple good answers for the same question.


\subsection{Logistic Regression}
\label{sec:logistic}

We use logistic regression to understand which factors drive user actions on Stack Exchange. 
Because our data is highly multi-dimensional, and some attributes are strongly correlated with others, we use LASSO penalized regression, where parameters are determined by maximizing the likelihood function with the addition of a penalty to avoid overfitting \cite{LASSO}. The value of this penalty was adjusted such that the mean squared error from 10-fold cross-validation was minimzed. As a check, we did the same fits with a different type of penalty, ridge regression, and found the behavior to be qualitatively the same. The fitting was performed with the R package ``glmnet'' \cite{glmnet}, which allows for fast and accurate determination of regression coefficients, \boldmath$\beta$.

We checked the robustness of our results by omitting data from the largest community for each board type (meta, non-technical, and technical), and re-determining the regression parameters.
The qualitative results were unaffected, and quantitatively, the results were very similar. For the rest of the paper, we focus on LASSO penalized regressions with all boards included.

\subsection{Deviance Ratio}

We use deviance ratio to determine how well the model fits the data.
The deviance ratio is reminescent of $R^2$, although it is used for models that maximize the likelihood function rather than minimize the mean squared error. 

The deviance ratio is defined as:
\begin{equation}
R_{dev} = 1 - \frac{D_{fit}}{D_{null}},
\end{equation}

where
\begin{equation}
D = -2 \left\{\text{log}[p(y| \hat\theta_0)] - \text{log}[p(y| \hat\theta_s)] \right\}
\end{equation}

In this case, $\text{log}[p(y| \hat\theta_s)]$ is the log-likelihood of the saturated model, with one degree of freedom per observation, while $\text{log}[p(y| \hat\theta_0)]$ is the log-likelihood for the fitted model. $D_{fit}$ is for the best fit model, while $D_{null}$ is the null intercept model. A careful observation reveals $D$ is simply $-2 \times$ (the log likelihood ratio), therefore the deviance ratio tells us how much of the likelihood ratio for the null model can be explained with a fitted model. Errors for this value are defined in the next section.

\subsection{Error}

The uncertainty in \boldmath$\beta$ and the deviance ratio (shaded regions in subsequent figures) is defined as the range of values such that, by changing the LASSO regression bias, the mean 10-fold cross-validated error (in this case, the deviance) is within one standard deviation of the minimum mean cross-validated error.
This spread of values is the clearest way we are aware of to show parameter uncertainty or sensitivity, because LASSO regression, like all penalized regression methods, does not have a standard method to caluculate uncertainties with high dimensional data~\cite{Penalized}.

\subsection{Attributes and Normalization}
\label{sec:attributes}
We use the following answer attributes in analysis:
\vspace{-5pt}
\begin{itemize}
\setlength\itemsep{-0.3em}
\item answerer's \emph{reputation} at the time the answer was created,
\item mean rate of \emph{reputation increase} over time,
\item answer's Flesch Reading Ease~\cite{Readability}, or \emph{readability}, score,
\item answerer's \emph{tenure} (i.e., time since joining the site) at the time of the answer,
\item number of \emph{hyperlinks} per answer,
\item binary value denoting whether the \emph{answer was eventually accepted} (for voting only),
\item answer \emph{score} before each vote,
\item default \emph{web page order} for an answer (i.e., its relative position),
\item \emph{chronological order} of an answer (whether it was first, second, third, etc.),
\item \emph{time} since an answer was created, or its age
\item \emph{number of words} per answer,
\item answer's \emph{word share}, that is the fraction of total words in all answers to the question.
\end{itemize}

Answerer reputation~\cite{Shah10}, Flesch readability, and word count~\cite{Ponzanelli14} were used in previous works as measures of answer quality, and often a ``high quality'' answer was at least in part defined as the accepted answer~\cite{Shah10,Kim09,Agichtein08}. To adeqately compare datasets, we removed all data where the question was not eventually accepted within the collection timeframe. Qualitatively, voters and askers in unaccepted questions had similar behavior to those in accepted questions, but quantitatively, we found variations on regression coefficients, potentially suggesting voters behave differently in this hold-out set. We also consider an answer's rank in the list of answers (what we refer to as web page order) and score, because these variables affect how much attention the answer receives~\cite{Salganik06,lerman14as,PopularDynam}. The other attributes were also examined as additional factors that could affect how answers are voted or accepted. These were, however, not found to significantly affect the results.

There is large variability in attribute values within and across the attributes. To account for the variability, we \emph{normalize} all attributes by mapping them to their associated cumulative distribution function (CDF). CDF normalization is non-parametric and accounts for the distribution of attribute values. An advantage of this normalization is that outliers have a minimal effect because values are evenly spread and bounded between 0 and 1. Normalization allows us to compare the relative importance of different attributes by comparing their regression coefficients. For web page order attribute, we divided by the number of answers available, which is equivalent to a CDF for the number of answers equal to 2, 3, etc., while for all other attributes, we used the CDF across all answers on all Stack Exchange communities.

To verify the selected attributes, we checked each attribute to make sure correlations with other attributes were reasonably low, and, if they were greater than $0.7$, we checked whether removal of the attribute increased the cross-validated error significantly. This correlation condition seems very liberal, but we wanted to include as many attributes used in previous literature as possible, and then use penalized regression to appropriately reduce the effect of colinearity. To check if this affected our results, we separately removed wordshare, score, whether the answer was eventially accepted, and webpage order, and found results were qualitatively the same.



\section{Results}
\label{Prediction Tasks}

%
%
%
%
%
%
%
%
%
%

We analyze Stack Exchange data to understand what attributes are strongly associated with the decision to vote for, or accept, an answer.
To do this, we find all attribute values just before an answer was voted for (or accepted), and then estimate attribute coefficients for a logistic regression model.

\subsection{Taming Heterogeneity}

Automatically uncovering homogeneous populations within heterogeneous observational data remains an open research challenge. In our study of Stack Exchange, we used exploratory data analysis to identify potential sources of heterogeneity. For example, users who are interested in technical topics (e.g., programming) may be driven by different factors to contribute to Stack Exchange than those who are interested in non-technical subjects (e.g., cooking), or governance (meta boards). To account for this source of heterogeneity, we split the data by the type of board---technical, non-technical and meta---and run regression analysis separately on each dataset.
We further split data by whether the asker eventually accepted an answer in our observation window, how an answer is chosen (vote versus accept), and the number of answers, but find that the greatest source of heterogeneity is the number of answers a question has at the time the user votes for or accepts it.

In the next two sections, we discuss our findings in greater detail, including the implications of the most important attributes, and the reasons for the heterogeneity in our data. Regression fits suggest users who vote when many answers are visible strongly depend on a small set of heutistic-like attributes compared to users who vote when there are few. Furthermore, askers are found to depend on heurstics much more than voters, which undermines the assumption that accepted answers are probably one of the best answers \cite{Shah10,Kim09,Agichtein08}. Overall, we find evidence that the wisdom of crowds in Stack Exchange boards could be reduced by the number of answers to a question, and the role of the user.

\subsection{Answer Attributes and Behavior}

\begin{figure*}[tbh]
	\hspace{-12pt}
	\includegraphics[width=1\textwidth]{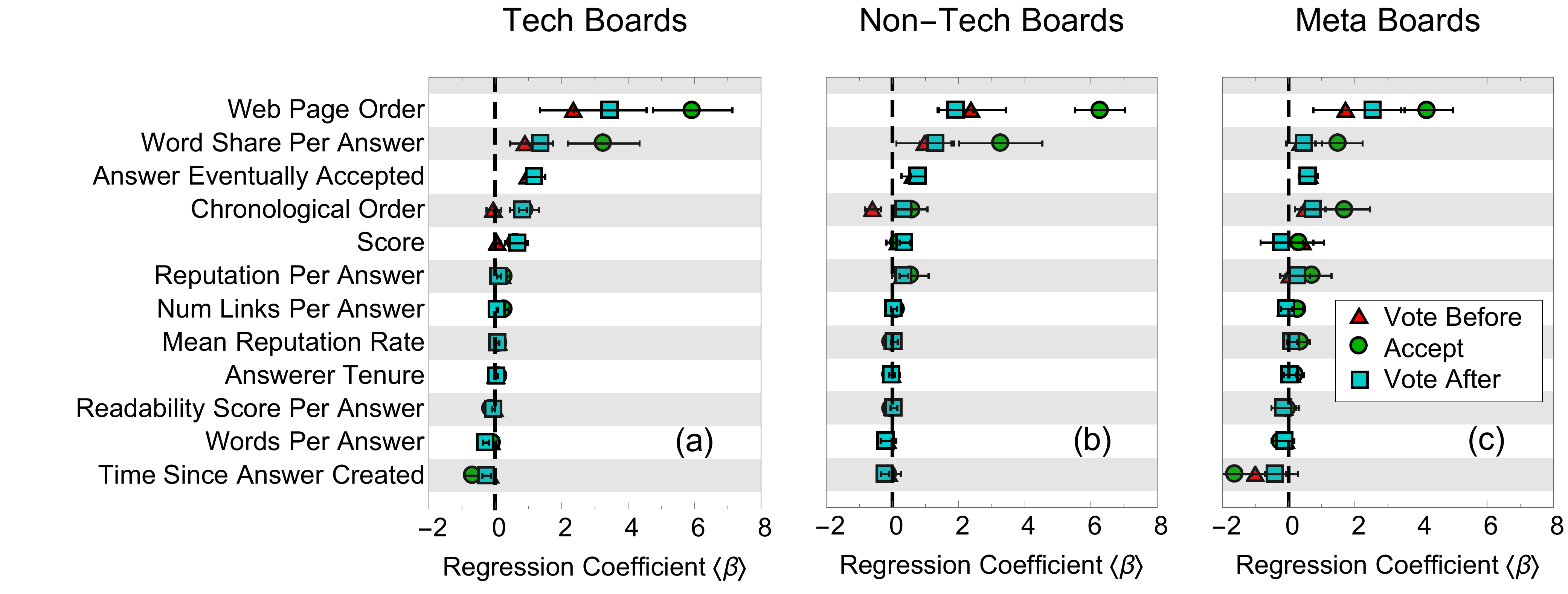}

	\caption{Regression coefficients for answerers to accept (green circles) and voters to vote for an answer both before (red triangles) and after (blue squares) an answer is accepted on  (a) technical, (b) non-technical, and (c) meta boards, averaged over the number of available answers from 2-20. Higher values indicate a stronger relationship between attributes and user behavior (voting or accepting an answer). Error bars indicate the variance of these values as the number of answers increases.
}
	\label{fig:ComparisonOfParams}
\end{figure*}

We take logistic regressions for votes cast before any answer was accepted, votes after an answer was accepted, as well as accepted answers. The average and variance of the regression parameters across $2-20$ answers are shown in Figure~\ref{fig:ComparisonOfParams}. Because all attributes were normalized, the larger the value, the more the respective parameter affects user behavior, relative to all others in the regression.

\begin{figure*}[tbh]
	\includegraphics[width=1\textwidth]{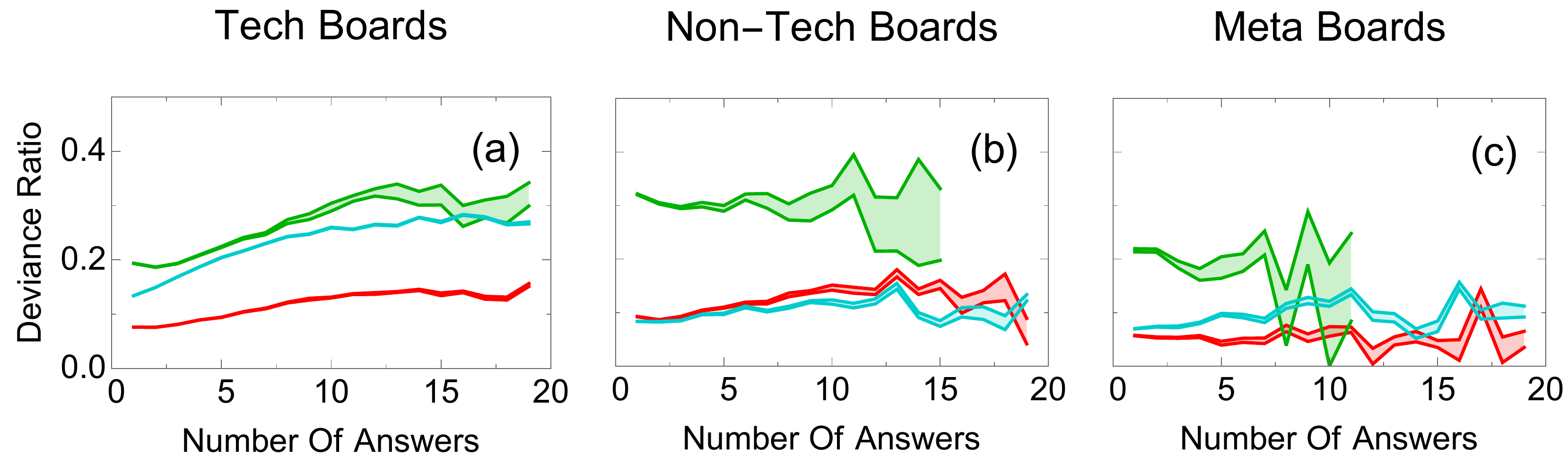}

	\caption{The deviance ratio (fraction of deviance explained by the model) for votes before acceptance (red), answer acceptance (green), and votes after acceptance (blue), for (a) technical, (b) non-technical, and (c) meta boards, with 2 to 20 answers. The shaded region represents the uncertainty in our values (see Section~\ref{Methods}). Askers have a larger deviance ratio, and therefore appear to be better modeled by our regressions, compared to answerers. Furthermore, the deviance ratio of voters tends to increase with the number of answers, suggesting increasing agreement with our model.
}
	\label{fig:DevRatio}
\end{figure*}

\begin{figure*}[tbh]
	\includegraphics[width=1\textwidth]{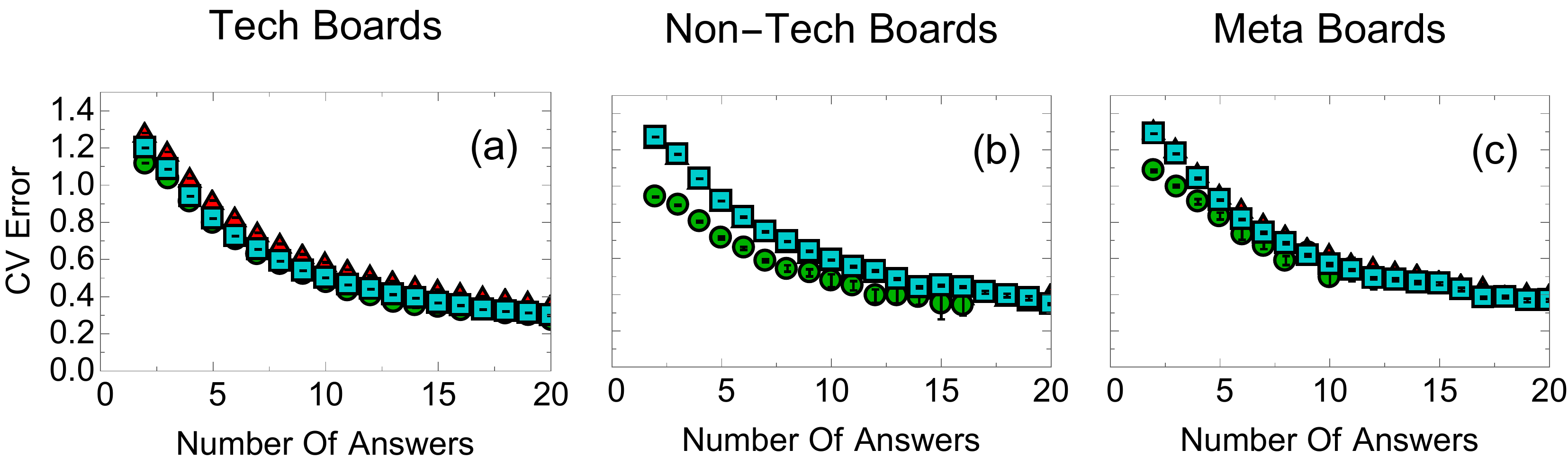}

	\caption{Mean deviance from 10-fold cross-validation for votes before acceptance (red), answer acceptance (green), and votes after acceptance (blue), for (a) technical, (b) non-technical, and (c) meta boards, with 2 to 20 answers. The deviance for votes before and after acceptance almost completely overlap. Askers have a lower prediction error compared to voters, but the most significant drop in deviance for all users occurs when the number of answers increases. 
}
	\label{fig:CVerror}
\end{figure*}

\begin{figure*}[tbh]

	\includegraphics[width=1\textwidth]{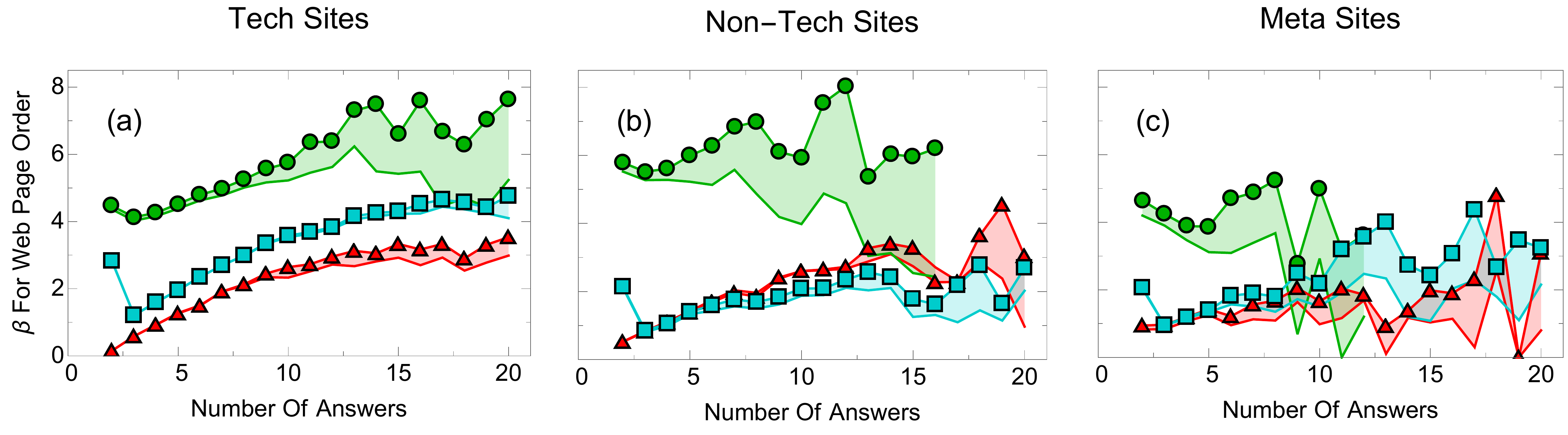}
	\caption{Web page order regression coefficients for voting before (red triangles) and after (blue squares) an answer is accepted, as well as accepting an answer (green circles) for (a) technical, (b) non-technical, and (c) meta boards, with 2 to 20 answers. The shaded region represents the uncertainty in our values (see Section~\ref{Methods}). Users increasingly depend on the web page order of an answer as the number of answers increases.
}
	\label{fig:WebPageOrder}
\end{figure*}

\begin{figure*}[tbh]
	\includegraphics[width=1\textwidth]{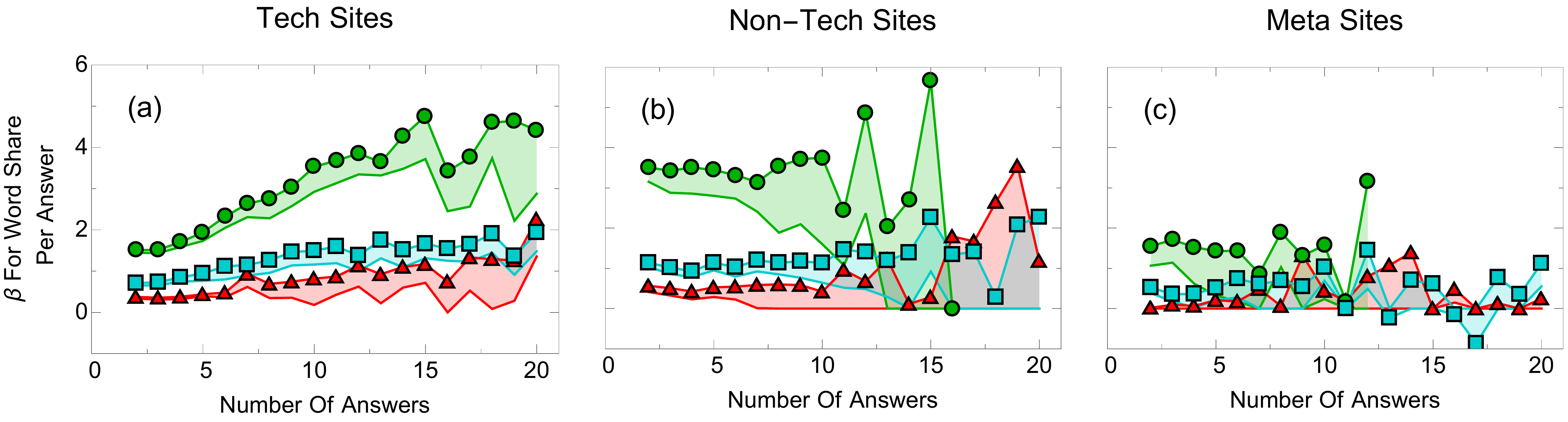}
	\caption{Word share regression coefficients for voting before (red triangles) and after (blue squares) an answer is accepted, as well as accepting an answer (green circles) for (a) technical, (b) non-technical, and (c) meta boards, with 2 to 20 answers. The shaded region represents the uncertainty in our values (see Section~\ref{Methods}).
Across all baords, voters appear increasingly likely to choose answers that take up a relatively large amount of web page space as the number of answers grows.
}
	\label{fig:WordShare}
\end{figure*}

We find that
web page order and word share are the two most important factors for users to choose an answer (Fig.~\ref{fig:ComparisonOfParams}). Because strong correlations may affect the coefficient of a particular attribute in penalized regression, we also remove each attribute separately (not shown), and find the cross-validated (CV) error decreases the most when the highest-coefficient attributes are removed, thus validating the usefulness of CDF normalization.

These findings alone are not necessarily surprising. We know from previous research that people's choices are biased by the rank order of items~\cite{Craswell08,lerman14as,PopularDynam}. Word share is potentially correlated with higher answer quality, because relatively long answers may be more informative, or they may just be easier to see (take up a large portion of the web page space).
We notice that both of these regression coefficients are even higher for askers than voters, across different board types, already suggesting a surprising degree of heterogeneity. 
Other factors, however, such as an answerer's reputation or tenure, how thoroughly an answer is documented with hyperlinks, how easy it is to read (readability), etc., do not seem to play a big role in users' choices of which answers to vote or accept.

\subsection{Behavior vs Number of Answers}
What is more surprising than the overall size of the regression coefficients, however, is that the largest coefficients, e.g., for web page order and word share, change substantially as the number of available answers to a question increases (Fig.~\ref{fig:WebPageOrder} and Fig.~\ref{fig:WordShare}). Furthermore, the models describe the data increasingly well (Fig.~\ref{fig:DevRatio}) and reduce predictive error (Fig.~\ref{fig:CVerror}). In other words, users' future decisions appear to be increasingly dependent on these attributes. This is also seen when we remove each attribute and check the resulting CV error of the model (not shown). We find that removing attributes, such as whether the answer was accepted or its web page order, would increasingly impact the CV error of voters as the number of answers grow.

\begin{figure*}[tbh]
\includegraphics[width=1\textwidth]{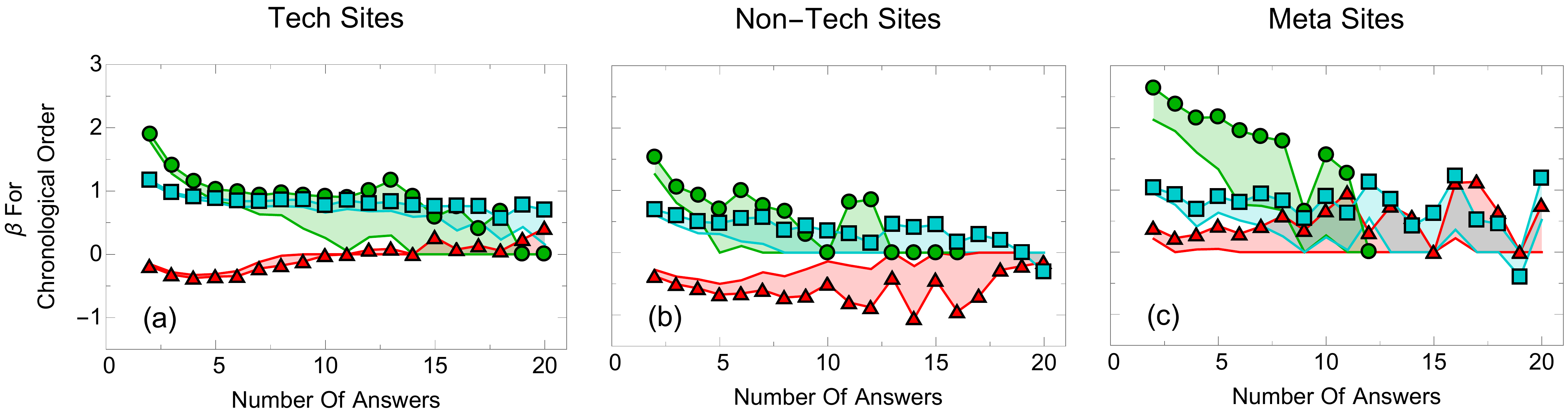}
	\caption{Chronological answer order regression coefficients for voting before (red triangles) and after (blue squares) an answer is accepted, as well as accepting an answer (green circles) for (a) technical, (b) non-technical, and (c) meta boards, with 2 to 20 answers. The shaded region represents the uncertainty in our values (see Section~\ref{Methods}). 
For all boards, there is a decreasingly significant dependence on the order in which answers appear. For askers and voters after acceptance, newer answers are preferred, while, for voters before acceptance, older answers are preferred.
}
	\label{fig:ChronologicalOrder}
\end{figure*}

A number of plausible explanations exist:
\begin{itemize}
\item The subsequent answers improve upon the previous answer, or
\item Some unknown confounding variable affects both the number of answers as well as user behavior, or finally
\item User behavior changes as a function of the number of available answers.
\end{itemize}

According to the first hypothesis, the last answer may be such an improvement on the previous ones that users will ``flock'' to it. Therefore, it should be no surprise that as the number of answers increases, changes in votes are seen. In theory, this should be captured by a significant dependence on answer's chronological order: voters should prefer newer answers to older ones.
In practice, this does not seem to be the case. The dependence on chronological order is relatively small (Figure~\ref{fig:ComparisonOfParams}), and furthermore decreases with the number of answers (Figure~\ref{fig:ChronologicalOrder}), which is exactly the opposite of what should be expected if this hypothesis were true.


The second hypothesis says that the number of answers and the behavior of the user both correlate to something else entirely; the results presented so far could be strongly affected by some confounding variable. For example, \cite{Anderson12} finds that the reputation of later answerers on Stack Overflow, a technical board within Stack Exchange devoted to programming questions, is lower than the reputation of earlier answerers. If later  voters similarly differ in reputation or some other attribute, this could potentially explain our results. We call this the ``lazy voter" hypothesis, because later voters may simply be ``lazier" and rely on heuristics to a greater extent. It is curious, however, that voter behavior does not seem to be significantly affected by the age of the answer, based on our regressions, and instead on the shear number of answers, as time progresses.

The last hypothesis is that users behave differently as the number of answers grows.
Economics and psychologists believe that people usually do not have the time, nor inclination or cognitive resources, to process all available information, but instead, employ heuristics to quickly decide what information is important. This phenomenon, known as {``bounded rationality''}~\cite{Kahneman03, simon1982models}, profoundly affects what information people pay attention to and the decisions they make~\cite{Kahneman11}. Our results suggest that rather than thoroughly evaluating all available answers to a question on Stack Exchange, users employ cognitive heuristics to choose the ``best'' answer. These heuristics include choosing top-ranked  answer (Fig.~\ref{fig:WebPageOrder}) or one that occupies more screen space (Fig.~\ref{fig:WordShare}).
These heuristics become more pronounced when the volume of information (number of available answers) grows.

Instead of being a cognitive heuristic, word share could plausibly reflect answer quality: high quality answers may be wordy. Interestingly, however, the regression coefficient for the number of words for each answer (rather than its share of words) is slightly negative, suggesting users overall prefer somewhat shorter answers if they prefer anything at all. It is intuitive that longer answers are more salient and catch a user's eye, especially when there are many answers.

Whether second or third hypothesis is true, our observation of a strong dependence of votes and accepts on the number of available answers suggests a strong limitation of crowdsourcing answer quality: collective judgment of quality may change with the number of answers, which is especially noticable with popular, and presumably important, questions which have many answers available (Fig.~\ref{fig:NumbAnswers}).

\begin{figure*}[tbh]
 		\includegraphics[width=1\textwidth]{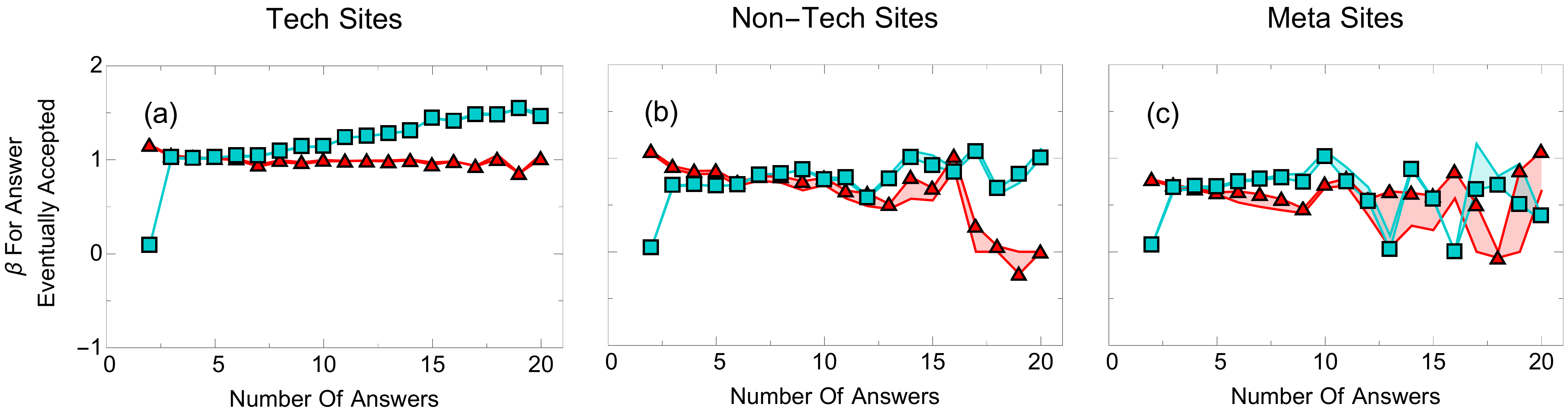}
	\caption{Regression coefficients for voting on an (eventually) accepted answer before (red triangles) and after (blue squares) that answer is accepted for (a) technical, (b) non-technical, and (c) meta boards, with 2 to 20 answers. The shaded region represents the uncertainty in our values (see Section~\ref{Methods}). There is a large and increasing vote dependence on the accepted answer once the asker accepts it, compared to before the answer is accepted, meaning the signal that this answer is accepted appears to have a statistically significant effect on voter behavior.
}
	\label{fig:AnswerAccepted}
\end{figure*}

We see further evidence of the final two arguments in Figure~\ref{fig:AnswerAccepted}, where we plot the regression coefficients for accepting an answer as a function of number of answers for voters before, and after, an answer is accepted.

We find that voters are more likely to choose an answer that is eventually accepted (the regression coefficients are positive), but, curiously, voters are even more likely to choose the answer \emph{after} it is accepted as the number of answers increase (the regression coefficient is usually even higher, and increases with the number of answers). 
In other words, although answer quality does not change before or after acceptance, users are more likely to vote on whatever the asker chooses, especially as the number of answers increases. This could either be due to ``lazy voters", who appear later on, when the number of answers is high, or because voters are overwhelmed by the number of answers.


Finally, askers are much better modeled by our regressions compared to voters (Fig.~\ref{fig:DevRatio}) and similarly, are more predictable (Fig.~\ref{fig:CVerror}). 
To better understand what we are seeing, we must understand Stack Exchange's rules. Namely, voters need a reputation above 15 in order to vote, which becomes a barrier to entry: typically, users must have provided answers or questions in the past that others upvoted in order to be able to vote. Askers on the other hand require less reputation. Presumably they rely more on heuristics than voters, because they are less able to recognize the correct answer. This is important because accepted answers have been used as a gold standard of answer quality in previous research \cite{Shah10,Kim09,Agichtein08}, but, if askers strongly rely on heuristics like answer rank order, this puts into question whether accepted answers are the best standard. Instead, we find that highly voted answers may be a better standard, because voters appear to depend less on heuristics. The correlation between score and answer acceptance is surprisingly low ($0.1-0.2$), so we have a strong incentive to explore in the future whether the highest scoring answer is a more effective quality standard.

\section{Conclusion}

We analyzed user activity over a five year period on 250 \qa~communities on the Stack Exchange network.
The goal of our study was to understand what factors influence users to vote for, or accept, particular answers. Analysis from our models of voter and asker behavior suggest that Stack Exchange users rely on simple cognitive heuristics to choose an answer to vote for or accept, especially as the number of answers available increases. First, model parameters describing the dependence of behavior on answer's web page order and word share increase with the number of available answers. Such dependence would not necessarily exist if web page order and word share were merely proxies for answer quality. Second, askers appear to rely more on heuristics compared to voters, who need higher reputation and therefore may be more proficient Stack Exchange users. This suggests that answer acceptance might not be the best proxy for answer quality. Finally, voters are more likely to vote for an answer after it is accepted than before that very same answer is accepted as the number of available answers grow. Not only does acceptance appear to change a user's judgment of answer quality, it appears to become an increasingly strong bias with the number of answers.

The behaviors we describe are consistent with, but not proof of, bounded rationality, in which decision-makers employ cognitive heuristics to make quick decisions instead of evaluating all available information~\cite{Simon71,Kahneman11}. Moreover, people tend to use heuristics to cope with the ``cognitive strain'' of information overload~\cite{Taylor1975psychological}. Psychologists and behavioral scientists have identified a wide array of cognitive heuristics, which introduce predictable biases into human behavior. Social influence, \emph{aka} ``bandwagon effect'', is one such heuristic: people pay attention to the choices of others~\cite{Salganik06}. We find, however, that this affect is not very significant in Stack Exchange. Another important heuristic for online activity is ``position bias''~\cite{Payne51}: people pay more attention to items at the top of the list or the screen than those below~\cite{lerman14as}. Position bias, or rank order, plays a large effect in user choices even after accounting for item quality~\cite{Hogg2015hcomp,lerman14as}, which is in agreement with the results presented here.
Alternative explanations of our results, however, are plausible. For example ``lazy" (more heuristically driven) voters might arrive later, after a question has many answers. 

No matter which explanation holds, however, our work offers a cautionary note to designers of crowdsourcing systems, such as Stack Exchange: collective judgments about content quality are not necessarily accurate.
To partly address this problem, the order in which answers are presented to users could be randomized, or questions could be closed to voting after some time.


Our work makes a number of methodological contributions valuable to the Data Science community. First, we use CDF normalization to make all variables commensurate. While this is a nonlinear transformation, it accounts for the distribution of variable values in the dataset, which reduces the influence of outliers and allows for fair comparison of heterogeneous variables.  Also, we handled behavioral heterogeneity by splitting by board type and number of answers. To check robustness of regression results, we used two types of penalized regression and ``leave out the largest board" analysis. These methods can be applied to model other heterogeneous behavioral data. Finally, we measured the uncertainty in parameter coefficients as the range of coefficients, due to varying the penalization in our regressions, such that the CV error is within one standard deviation of the minimum mean error. We are unaware of alternative methods to accurately display the uncertainty in measurements from penalized regressions, and the uncertainties from our regressions appear reasonable.


Our analysis of observational data cannot completely control for the some of the known (and unknown) covariates that can affect our conclusions. For example, we cannot completely separate the effects of cognitive heuristics from those of answer quality.
A necessary step in future research is to conduct a laboratory study to control for variation in answer quality, similar to previous studies~\cite{Hogg2015hcomp,lerman14as}, to quantify the degree to which crowds are ``myopic.'' Despite known limitations, our work highlights the benefits of using data mining to understand and predict human behaviors, and may provide insight into improving the quality and performance of crowdsourcing systems.

\label{lastpage}

\begin{thebibliography}{10}

\bibitem{adamic}
L.~A. Adamic, J.~Zhang, E.~Bakshy, and M.~S. Ackerman.
\newblock Knowledge sharing and yahoo answers: everyone knows something.
\newblock In {\em WWW}, pp 665--674. 2008.

\bibitem{Agichtein08}
E.~Agichtein, C.~Castillo, D.~Donato, A.~Gionis, and G.~Mishne.
\newblock Finding high-quality content in social media.
\newblock In {\em WSDM}, pp 183--194. 2008.

\bibitem{agosto}
D.~E. Agosto.
\newblock Bounded rationality and satisficing in young people's web-based
  decision making.
\newblock {\em J. American Society for Information Science and
  Technology}, 53(1):16--27, 2002.

\bibitem{Anderson12}
A.~Anderson, D.~Huttenlocher, J.~Kleinberg, and J.~Leskovec.
\newblock Discovering value from community activity on focused question
  answering sites: a case study of stack overflow.
\newblock In {\em KDD}, pp 850--858. 2012.

\bibitem{blooma}
M.~J. Blooma, D.~Hoe-Lian~Goh, and A.~Yeow-Kuan~Chua.
\newblock Predictors of high-quality answers.
\newblock {\em Online Information Review}, 36(3):383--400, 2012.

\bibitem{chen}
B.-C. Chen, A.~Dasgupta, X.~Wang, and J.~Yang.
\newblock Vote calibration in community question-answering systems.
\newblock In {\em SIGIR}, pp 781--790. 2012.

\bibitem{Chen14}
H.~Chen, P.~De, Y.~J. Hu, and B.-H. Hwang.
\newblock Wisdom of crowds: The value of stock opinions transmitted through
  social media.
\newblock {\em Rev. Financ. Stud.}, 27(5):1367--1403, 2014.

\bibitem{Craswell08}
N.~Craswell, O.~Zoeter, M.~Taylor, and B.~Ramsey.
\newblock An experimental comparison of click position-bias models.
\newblock In {\em WSDM}, pp 87--94, 2008.

\bibitem{glmnet}
J.~Friedman, T.~Hastie, and R.~Tibshirani.
\newblock Regularization paths for generalized linear models via coordinate
  descent.
\newblock {\em J Stat Softw.}, 33:1--22, 2010.

\bibitem{ghosh2014game}
A.~Ghosh and P.~Hummel.
\newblock A game-theoretic analysis of rank-order mechanisms for user-generated
  content.
\newblock {\em J. Economic Theory}, 154:349--374, 2014.

\bibitem{Penalized}
J.~J. Goeman.
\newblock L1 penalized estimation in the cox proportional hazards model.
\newblock {\em Biometrical Journal}, 52(1):70--84, 2010.

\bibitem{LASSO}
T.~Hastie, R.~Tibshirani, and J.~Friedman.
\newblock {\em The Elements of Statistical Learning: Data Mining, Inference,
  and Prediction.}
\newblock Springer, 2009.

\bibitem{Hodas14srep}
N.~O. Hodas and K.~Lerman.
\newblock The simple rules of social contagion.
\newblock {\em Scientific Reports}, 4, 2014.

\bibitem{Hogg2015hcomp}
T.~Hogg and K.~Lerman.
\newblock Disentangling the effects of social signals.
\newblock {\em Human Computation Journal}, 2(2):189--208, 2015.

\bibitem{jain2009designing}
S.~Jain, Y.~Chen, and D.~C. Parkes.
\newblock Designing incentives for online question and answer forums.
\newblock In {\em EC}, pp 129--138. 2009.

\bibitem{Kahneman03}
D.~Kahneman.
\newblock Maps of bounded rationality: Psychology for behavioral economics.
\newblock {\em American Economic Review}, 93(5):1449--1475, 2003.

\bibitem{Kahneman11}
D.~Kahneman.
\newblock {\em Thinking, fast and slow}.
\newblock Farrar, Straus and Giroux, 2011.

\bibitem{Kim09}
S.~Kim and S.~Oh.
\newblock Users relevance criteria for evaluating answers in a social q\&a
  site.
\newblock {\em J. Assoc. Inf. Sci. Techno.}, 60(4):716--727, 2009.

\bibitem{Readability}
J.~P. Kincaid, R.~P. Fishburne, Jr, R.~L. Rogers, and B.~S. Chissom.
\newblock Derivation of new readability formulas (automated readability index,
  fog count and flesch reading ease formula) for navy enlisted personnel.
\newblock Technical report, U.S. Navy, 1975.

\bibitem{Wagner10}
A.~Kittur and R.~Kraut.
\newblock The wisdom of reluctant crowds.
\newblock In {\em HICSS}, pp 1--10, Honolulu, HI, 2010. IEEE.

\bibitem{lerman14as}
K.~Lerman and T.~Hogg.
\newblock Leveraging position bias to improve peer recommendation.
\newblock {\em PLoS ONE}, 9(6):e98914, 2014.

\bibitem{Lorenz11}
J.~Lorenz, H.~Rauhut, F.~Schweitzer, and D.~Helbing.
\newblock How social influence can undermine the wisdom of crowd effect.
\newblock {\em PNAS},
  108(22):9020--9025, May 2011.

\bibitem{Malhotra84}
N.~K. Malhotra.
\newblock Reflections on the information overload paradigm in consumer decision
  making.
\newblock {\em J. Consum. Res.}, pp 436--440, 1984.

\bibitem{Payne51}
S.~L. Payne.
\newblock {\em The Art of Asking Questions}.
\newblock Princeton University Press, 1951.

\bibitem{Ponzanelli14}
L.~Ponzanelli, A.~Mocci, A.~Bacchelli, M.~Lanza, and D.~Fullerton.
\newblock Improving low quality stack overflow post detection.
\newblock In {\em ICSME}, pp 541--544. IEEE, 2014.

\bibitem{rechavi}
A.~Rechavi and S.~Rafaeli.
\newblock Not all is gold that glitters: Response time \& satisfaction rates in
  yahoo! answers.
\newblock In {\em SocialCom}, pp 904--909. IEEE, 2011.

\bibitem{Rodriguez14}
M.~G. Rodriguez, K.~Gummadi, and B.~Schoelkopf.
\newblock Quantifying information overload in social media and its impact on
  social contagions.
\newblock In {\em ICWSM}, Mar. 2014.

\bibitem{sakai}
T.~Sakai and R.~Song.
\newblock Evaluating diversified search results using per-intent graded
  relevance.
\newblock In {\em SIGIR}, pp 1043--1052. 2011.

\bibitem{Salganik06}
M.~J. Salganik, P.~S. Dodds, and D.~J. Watts.
\newblock Experimental study of inequality and unpredictability in an
  artificial cultural market.
\newblock {\em Science}, 311(5762):854--856, 2006.

\bibitem{scheibehenne2010can}
B.~Scheibehenne, R.~Greifeneder, and P.~M. Todd.
\newblock Can there ever be too many options? a meta-analytic review of choice
  overload.
\newblock {\em J. Consumer Research}, 37(3):409--425, 2010.

\bibitem{shah}
C.~Shah.
\newblock Measuring effectiveness and user satisfaction in yahoo! answers.
\newblock {\em First Monday}, 16(2), 2011.

\bibitem{Shah10}
C.~Shah and J.~Pomerantz.
\newblock Evaluating and predicting answer quality in community qa.
\newblock In {\em SIGIR}, pp 411--418.  2010.

\bibitem{Simon71}
H.~A. Simon.
\newblock Designing organizations for an information rich world.
\newblock In M.~Greenberger, editor, {\em Computers, communications, and the
  public interest}, pp 37--72. Baltimore, 1971.

\bibitem{simon1982models}
H.~A. Simon.
\newblock {\em Models of bounded rationality: Empirically grounded economic
  reason}, volume~3.
\newblock MIT press, 1982.

\bibitem{PopularDynam}
G.~Stoddard.
\newblock Popularity dynamics and intrinsic quality in reddit and hacker news.
\newblock In {\em ICWSM}, pp 416--425, 2015.

\bibitem{surowiecki2005wisdom}
J.~Surowiecki.
\newblock {\em The wisdom of crowds}.
\newblock Anchor, 2005.

\bibitem{Taylor1975psychological}
R.~N. Taylor.
\newblock Psychological determinants of bounded rationality: Implications for
  decision-making strategies.
\newblock {\em Decision Sciences}, 6(3):409--429, 1975.

\bibitem{BoundedRationality}
P.~Todd.
\newblock How much information do we need?
\newblock {\em European Journal of Operational Research}, 177(3):1317--1332,
  2005.

\bibitem{Vaupel85heterogeneity}
J.~W. Vaupel and A.~I. Yashin.
\newblock Heterogeneity's ruses: some surprising effects of selection on
  population dynamics.
\newblock {\em The American Statistician}, 39(3):176--185, 1985.

\bibitem{wang}
X.-J. Wang, X.~Tu, D.~Feng, and L.~Zhang.
\newblock Ranking community answers by modeling question-answer relationships
  via analogical reasoning.
\newblock In {\em SIGIR}, pp 179--186.  2009.

\end{thebibliography}
\end{document}